\pdfoutput=1 
\documentclass{collective-intelligence}

\usepackage{harvard}

\begin{document}
\bibliographystyle{agsm}

\title{MOTIVATIONS FOR PARTICIPATION IN SOCIALLY NETWORKED\\COLLECTIVE INTELLIGENCE SYSTEMS}
%
%
%
%
%

\numberofauthors{1} 
%
\author{
%
%
\alignauthor
Jon Chamberlain, Udo Kruschwitz, Massimo Poesio\\
       \affaddr{School of Computer Science and Electronic Engineering}\\
       \affaddr{University of Essex}\\
       \affaddr{Wivenhoe Park, Colchester CO4 3SQ, England}\\
       \email{\{jchamb,udo,poesio\}@essex.ac.uk}
}

\maketitle
\begin{abstract}
One of the most significant challenges facing systems of collective intelligence is how to encourage  participation on the scale required to produce high quality data. 
This paper details ongoing work with {\it Phrase Detectives}, an online game-with-a-purpose deployed on Facebook, and investigates user motivations for participation in social network gaming where the wisdom of crowds produces useful data.  

\end{abstract}

\section{Introduction}

Recent advances in human language technology have been made possible by acquiring large-scale resources created by novel methods of harnessing collective intelligence over the Internet. This approach is motivated by the 
observation that a group of individuals can contribute to a collective solution,
which has a better performance and is more robust than an individual's
solution. This is demonstrated in simulations of collective behaviour in
self-organizing systems \cite{Johnson98Symbiotic}.

Collective Intelligence (CI) systems such as Wikipedia and similar large 
initiatives have shown that a surprising number of individuals can be 
willing to participate in projects. 

A novel approach was the development of games-with-a-purpose (GWAP) that aggregate  data from non-expert players for collective decisions similar to what might be expected from an expert. 

Encouraging participation in GWAP projects remains a significant challenge, despite successful early efforts \cite{vonAhn06Games}. 

This paper investigates user participation in CI systems and whether social network platforms could offer anything to this approach. The paper then goes on to describe {\it Phrase Detectives}, a GWAP for creating
annotated language resources, specifically looking at the interface that was developed for the Facebook platform and the  modifications that were made to maximise the social incentives for players. Data from over a year of the game being live on Facebook is analysed and, in conclusion, suggestions are proposed for developers considering using a social networking platform for deployment of CI interfaces.

\section{Collective Intelligence Systems}

Collective intelligence can be shown in many domains including Computer Science, Economics and Biology\footnote{http://scripts.mit.edu/$\sim$cci/HCI} but here we focus on coordinating collective action in computational systems of CI. Individual decisions made by the community are aggregated in an attempt to produce a high quality, collective decision comparable to an expert judgement. 

One important  goal of CI systems is to overcome the bottleneck in creating and maintaining resources that would normally have to be done by paid administrators. Examples include encyclopedia websites like Wikipedia, citizen science projects for common knowledge and games-with-a-purpose that collect metadata.

By collecting decisions from a large, distributed group of contributors it is possible to approximate a single expert's judgements \cite{Snow(2008)}. This is in line with findings in other contexts \cite{Feng(2009),AlBakour10Sentence}.

\subsection{Participation}
The willingness of Web users to collaborate in  the
creation of resources is clearly illustrated by Wikipedia.
English Wikipedia numbers (as of October 2011) 3,773,941 articles,
written by over 15.5 million collaborators and 5,559 reviewers.\footnote{http://meta.wikimedia.org/wiki/List\_of\_Wikipedias}

\begin{figure*}
\centering
\scalebox{0.35}{\includegraphics{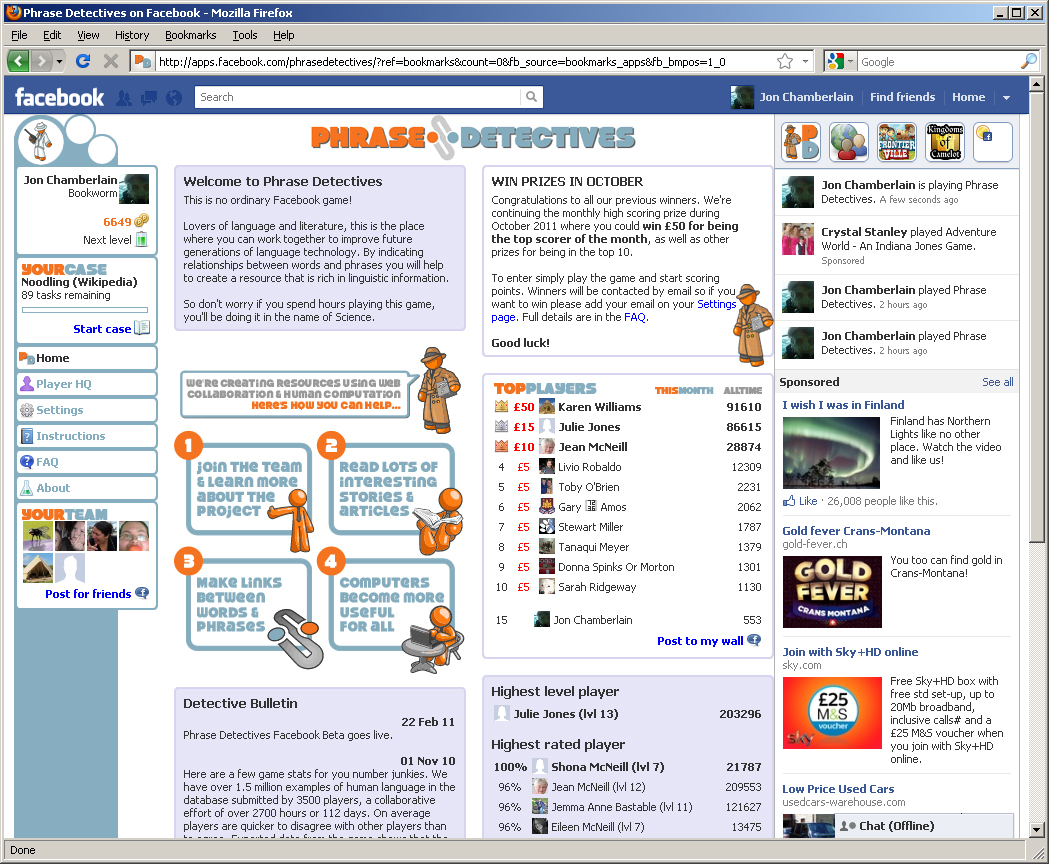}}
\caption{\label{PD2_home} Screenshot of the Phrase Detectives Facebook homepage.}
\end{figure*}

Citizen science projects, where non-expert volunteers complete large-scale or computationally difficult tasks,  include Open Mind Commonsense\footnote{http://openmind.media.mit.edu} (now ConceptNet\footnote{http://conceptnet.media.mit.edu})
which demonstrated that Web collaboration can be relied on to create resources, with 14,500 volunteers contributing nearly 700,000
sentences \cite{Singh02Public}.

The first, and perhaps most successful, game-with-a-purpose was The ESP Game\footnote{http://www.gwap.com/gwap} which
attracted over 200,000 players who have
produced over  50 million labels \cite{vonAhn06Games}.

Clearly there is huge potential for the general public to become engaged in CI systems and collaborate in producing resources that would not be possible to achieve using traditional methods.

\subsection{Social Networking Platforms}

Given the social nature of CI it seems logical to deploy CI systems on platforms where the users are already networked. In recent years social networking has become the dominant pastime online. As much as 22\% of time online is spent on social networks like Facebook, MySpace, Bebo, Twitter and others. This is three times the amount of time spent emailing and 7 times the amount of time spent searching the Internet.\footnote{http://mashable.com/2010/08/02/stats-time-spent-online}

The success of social network games such as Cityville, with over 50 million active players each month, or The Sims, Farmville and Texas HoldEm Poker, with over 30 million active monthly players each, show that the potential for large scale participation is possible using social networking platforms.\footnote{http://www.appdata.com} 

An estimated 927 million hours are spent each month by Facebook users playing games\footnote{http://www.allfacebook.com/facebook-games-statistics-2010-09}, which is another indicator of the vast human resource available.

A study of US and UK social network users showed that Facebook was by far the most frequently used platform for social network gaming (used by 83\% of users, compared to MySpace, the next highest platform, at 24\%).\footnote{http://www.infosolutionsgroup.com/2010\_PopCap\_Social \_Gaming\_Research\_Results.pdf} However Google are planning to increase their market share of this valuable resource.\footnote{http://www.reuters.com/article/2011/08/12/us-google-games-idUSTRE77A66H20110812}

Human language technology games integrated into social networking sites such as Sentiment Quiz\footnote{http://apps.facebook.com/sentiment-quiz} on Facebook show that social interaction within a game environment does motivate players to participate \cite{Rafelsberger2009}.

It is becoming more apparent that CI interfaces should be linked to 
social networking sites like Facebook to achieve high 
visibility, to explore different ways players can collaborate and to exploit this enormous human resource. 

\begin{figure}
\centering
\scalebox{0.45}{\includegraphics{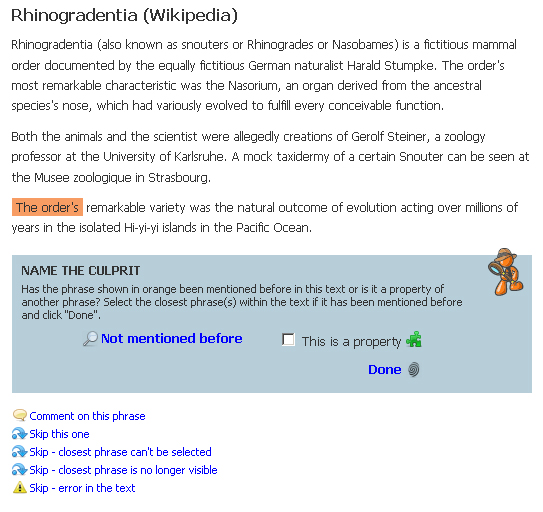}}
\caption{\label{PD2_ann} Detail of a task presented in Annotation Mode.}
\end{figure}

\section{Phrase Detectives Facebook Interface}

The first interface for the {\it Phrase Detectives} game\footnote{http://www.phrasedetectives.com}  \cite{Chamberlain08Phrase} enabled groups of players to work on the same task over a period of time as this was likely to lead to a collectively intelligent decision \cite{Surowiecki:05}. 

The Facebook version of {\it Phrase Detectives}\footnote{http://apps.facebook.com/phrasedetectives}, launched in February 2011, maintained the overall game architecture whilst incorporating a number of new features developed specifically for the social network platform (see Figure \ref{PD2_home}). 

The game was developed in PHP SDK (a Facebook API language allowing access to  user data, friend lists, wall posting etc) and integrates seamlessly within the Facebook site. Data generated from this version of the game is compatable with previous versions and both current implementations of the game run simultaneously on the same corpus of documents.

In order to play the game a Facebook user must grant certain permissions: the basic access (user details and friends list), which is required for all applications, and access to posting on the user's wall. Once the user has allowed the game access they never need to login to the game, only to Facebook.

\begin{figure}
\centering
\scalebox{0.45}{\includegraphics{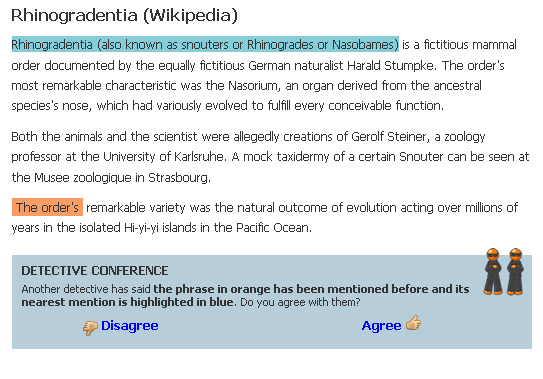}}
\caption{\label{PD2_val} Detail of a task presented in Validation Mode.}
\end{figure}

The game uses 2 styles of text annotation for players to complete a linguistic task. Initially text is presented in Annotation Mode (called Name the Culprit in the game - see Figure \ref{PD2_ann}). This is a straightforward annotation mode where the player makes an annotation decision about a highlighted markable (section of text). If different players enter different interpretations for a markable then each interpretation is presented to more players in Validation Mode (called Detectives Conference in the game - see Figure \ref{PD2_val}). The players in Validation Mode have to agree or disagree with the interpretation.

Players are trained with training texts created from a gold standard (a text that has been annotated by a linguistic annotation expert). A player always receives a training text when they first start the game and may also need to complete one when being promoted to the next level. Once the player has completed all of the training tasks they are given a rating (the percentage of correct decisions out of the total number of training tasks). The rating is recorded with every future annotation that the player makes as the rating is likely to change over time. 

\begin{figure}
\centering
\scalebox{0.60}{\includegraphics{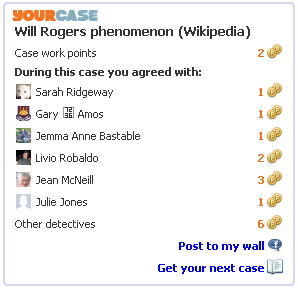}}
\caption{\label{PD2_casepoints} Detail of the reward screen, displayed at the end of each case, showing the player how many points they scored and who they agreed with.}
\end{figure}

The scoring system is designed to reward effort and motivate high quality decisions by awarding points for retrospective collaboration (see Figure \ref{PD2_casepoints}). 

The game makes full use of socially motivating factors inherent in the Facebook platform. Any of the player's friends who are playing the game form the player's team, which is visible in the left hand menu. Whenever a player's decision agrees with a team member they score additional points.

\begin{figure}
\centering
\scalebox{0.60}{\includegraphics{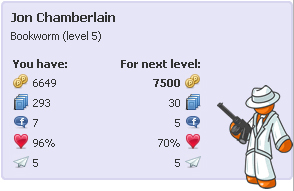}}
\caption{\label{PD2_levelcriteria}Detail showing criteria for the next level, displayed to the player on their homepage.}
\end{figure}

Player levels have well-defined criteria and the player must activate the new level once the criteria are met (see Figure \ref{PD2_levelcriteria}):

\begin{itemize} 
\item Total points scored
\item The player's rating
\item Documents completed
\item Training documents completed
\item Facebook posts made from the game
\end{itemize}

The Facebook game has monthly and all-time leaderboards as well as leaderboards for the highest level players, highest rated players and the players with the biggest team.

\section{Incentives}

There are 3 main incentive structures that can be used to encourage recruitment and participation: personal; social; and financial \cite{Chamberlain09New}. All incentives are applied with caution as rewards have been
known to decrease annotation quality
\cite{mrozinski-whittaker-furui:2008:ACLMain}. 

A previous survey showed that women are more likely to play social network games than men\footnotemark[9] and this could have an impact on the types of incentives offered. Facebook allows access to user data, including gender, as part of the basic access for the application so gender and workload of the players could be investigated.\footnote{It was assumed for the purposes of this investigation that Facebook users declare their gender truthfully.}

\subsection{Personal incentives}
Personal incentives are evident when simply participating is enough of a reward for the user. Generally, the most important personal incentive is that the user feels they are contributing to a worthwhile project. 

Also important for the players of {\it Phrase Detectives} is that they read
texts that they find interesting. The choice of documents is important in
getting users to participate in the game, to understand the tasks and to keep
playing. Whilst some texts are straightforward, others can provide a serious challenge of reading comprehension and completion of linguistic tasks. 

Texts were graded on complexity (on a scale of 1 to 4) on import. A player can choose the maximum level of document complexity they wish to read as they may be motivated to play the game to improve their English skills or, equally, because they enjoy reading challenging texts.

\subsection{Social incentives}
Social incentives reward users by improving their standing amongst their peers (in this case their fellow players and friends).

Using leaderboards and assigning levels for points can be an effective motivator, with players often using these as
targets i.e., they keep playing to reach a level or high score before stopping \cite{vonAhn08Designing}.

To investigate this in {\it Phrase Detectives}, players were grouped by how much progress they had made towards the next level in terms of points they had scored beyond the requirement for their current level. For example if a player had 110 points on level 2 (where the points requirement is 100) and the level 3 requirement is 200 points, then this player has progressed 10\% towards the next level.

\begin{figure}
\centering
\scalebox{0.40}{\includegraphics{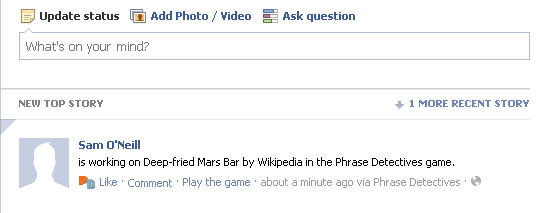}}
\caption{\label{PD2_wallpost} Detail of a news (or wall) post created automatically from the game, as seen by the player's friend.}
\end{figure}

News feed (or wall) posting is integrated into the game. This allows a player to make an automatically generated post to their news feed which will be seen by all of the player's friends (see Figure \ref{PD2_wallpost}).\footnote{Since the release of the game Facebook has changed how posts are displayed. Posts from the game now appear on the player's profile and in a news ticker.}

The posts include a link back to the game. Players are required to make a post from the game every time they are promoted to the next level. Posting is a very important factor in recruiting more players as studies have shown that the majority of social game players start to play because of a friend recommendation.\footnotemark[9]
\footnote{http://www.lightspeedresearch.com/press-releases/it's-game-on-for-facebook-users}

Posts may be social (display information about the document the player is working on or has just completed), collaborative (asking friends to join the game) or competitive (position in a leaderboard). Social posts are similar to information social network users share with friends so it is reasonable to assume they will be the most common type of post made from the game. This was investigated by analysing the logs of wall posts made from the game.

\subsection{Financial incentives}

Financial incentives reward effort with money. When financial rewards were offered in the game they were for monthly high scores, ranging from \textsterling 50 for first place, \textsterling 15 for second place, \textsterling 10 for third place and \textsterling 5 for fourth to tenth place. During July 2011 additional daily lottery-style prizes of \textsterling 5 were awarded, along with \textsterling 20 for the player with the highest level, highest rating and largest team. The monthly prizes motivate the high-scoring players to compete with each other by doing more work and motivate low-scoring players in the early parts of the month when the high score is low. The prizes were sent as Amazon vouchers by email to the winners. 

Whilst financial incentives are important to recruit new players, a combination of all 3 types of incentives is essential for the long term success of a game \cite{Smadja_2009}. 

The effectiveness of incentives was analysed by looking at new players, active players and new annotations each month. Months where prize funds were available were compared to months where there was none and a per-annotation cost effectiveness was calculated.  For the first 5 months no prize funds were offered but the following months all had prize funds of \textsterling 110, except July 2011 which had \textsterling 320 (see Table \ref{table}). 

The site was promoted in February, July and December 2011 to email lists, existing players, relevant Facebook groups etc, as well as advertised on GoogleAds and Facebook (a modest total pay-per-click budget of \textsterling 160) and competition websites. 

\section{Results}

Results from the Facebook interface of {\it Phrase Detectives} were analysed from February 2011 to February 2012.

\subsubsection{Gender and workload of players}

The Facebook game attracted 612 players of which 63\% were female, 35\% were male and 2\% did not disclose their gender. Of the ten highest scoring players, 60\% were female, 30\% were male and 10\% did not disclose their gender. This supports the previously mentioned survey that social network games are played predominately by women.

\begin{figure}
\centering
\scalebox{0.27}{\includegraphics{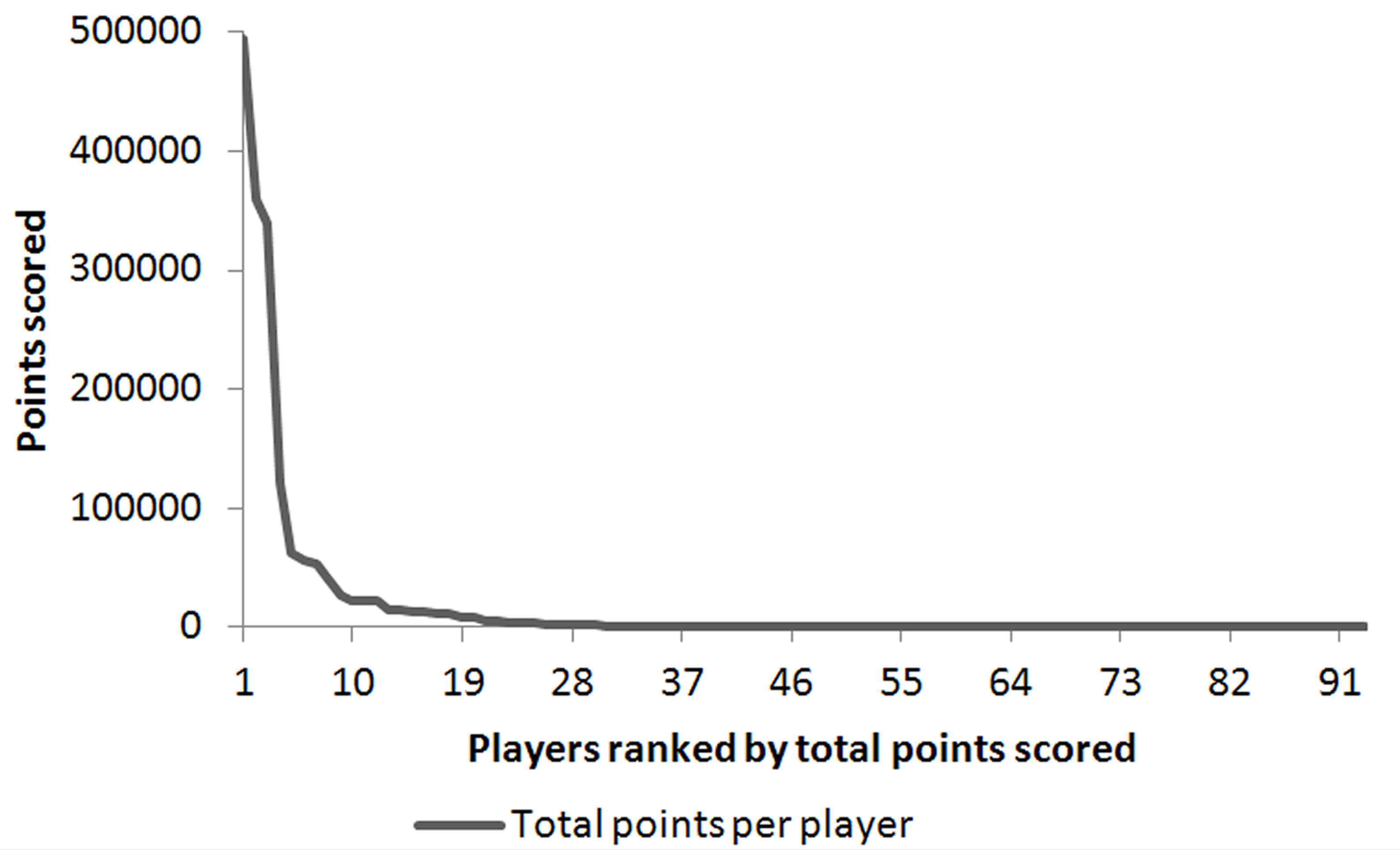}}
\caption{\label{R_players_points} Chart showing the scores of players.}
\end{figure}

\begin{figure}
\centering
\scalebox{0.27}{\includegraphics{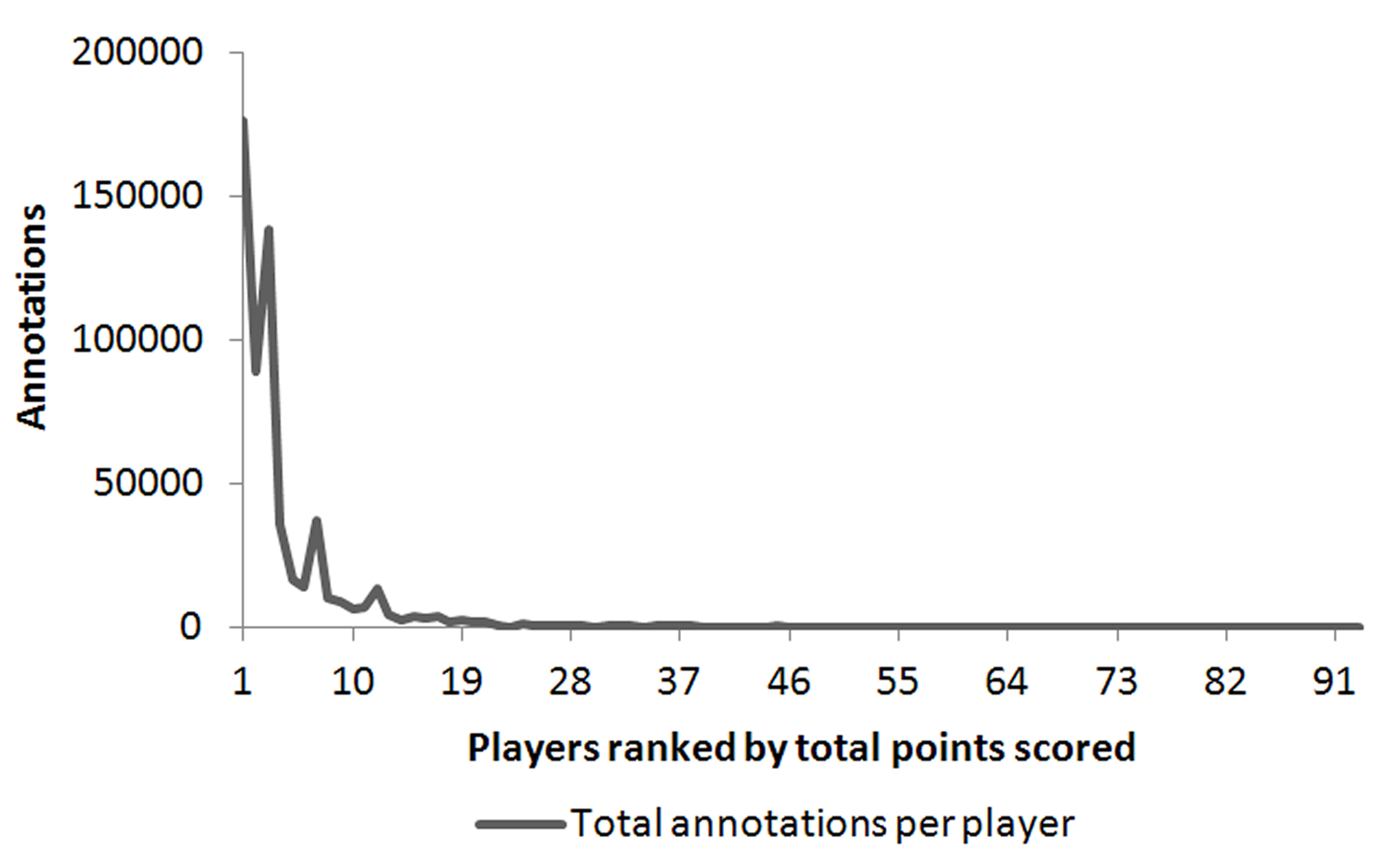}}
\caption{\label{R_players_av} Chart showing the total annotations of players, ranked by their total score.}
\end{figure}

In a study of the previous version of {\it Phrase Detectives} it was reported that the ten highest scoring players (representing 1.3\% of total players) had 60\% of the total points on the system and had made 73\% of the annotations\footnote{For the purpose of data analysis, annotations and validations are counted together and referred to as annotations.} \cite{Chamberlain09New}. In the Facebook version of the game the ten highest scoring players (representing 1.6\% of total players) had 89\% of the total points and had made 89\% of the annotations (see Figure \ref{R_players_points}).  The same ranking was used in Figure \ref{R_players_av} to show that although the total number of annotations of players is generally in line with points scored it is not always the case.

These results show that the majority of the workload (in this case annotation of documents) is being done by a handful of players. However, the influence of players who only contribute a little should not be undervalued as in some systems it can be as high as 30\% of the workload \cite{Kanefsky01} and this is what makes the collective decision making robust.

A subset of 112 players (those that declared their gender, had a rating above zero and had completed at least one annotation) were analysed to investigate whether gender was related to the amount of work a player does. Of these players men represented 35\% of the total (39 players). On average male players had completed 1,290 annotations and scored 4,636 points, compared to female players who had completed 4,636 annotations and scored 20,776 points. Although female players appear to be doing more work on average the difference is not statistically significant (using an unpaired t-test for significance testing). 

The ten highest scoring male players were then compared to the ten highest scoring female players. On average, the former  had completed 4,817 annotations and scored 17,628 points, whereas the latter had made 48,359 annotations and scored 144,905 points. The difference in both workload and score is statistically significant (p\textless 0.05).

This suggests that not only are female players more likely to play a socially networked game, they are also more likely to actively participate than male players. Further analysis is required to investigate whether female players provide higher quality annotations.





\begin{figure}
\centering
\scalebox{0.36}{\includegraphics{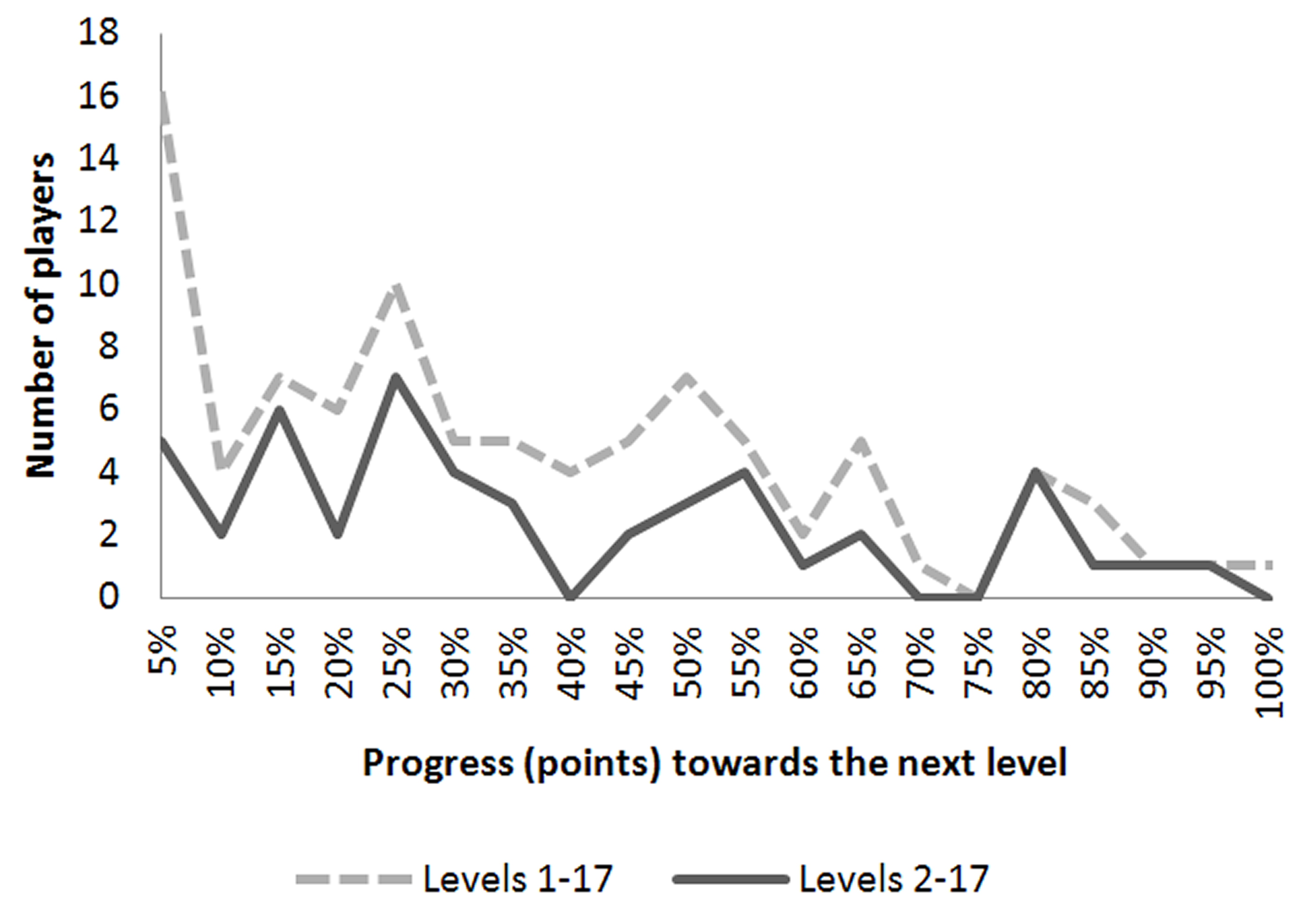}}
\caption{\label{R_scoreclustering} Chart showing the distribution of players' progress towards the next level.}
\end{figure}

\subsubsection{Using levels as goals}

All players who had a score (i.e. they had passed the training stage and had completed a document) were categorised in terms of their progress to the next level. In the first analysis all players from levels 1 to 17 were included. 

To distinguish between players who gave up after completing the training (i.e., the early stage of level 1) from the players stopping after reaching their goal of the next level, a second analysis was made excluding players on level 1.

The clustering of players near to the early stages of progress to the next level (see Figure \ref{R_scoreclustering}) could be an indication that players are motivated to work towards the next level before stopping, however the level criteria and scoring  system make this difficult to assess. Players score points in batches because all points scored on a document are added when the case is completed. Also, at this time, all extra points from other players who have agreed with their annotations are added. This effect becomes more negligible at higher levels where the progression to the next level is longer.

The analysis looks at player progression at the end of the data collection period i.e., players who get to the next level then never play the game again. To investigate this in detail the progression of each player session should be plotted however the game is not designed to test this and, if it were, many more players would be needed. 

Whilst it is intuitive to think that players will use level thresholds as goals this evidence does not support it.

\subsubsection{Posting from the game}

\begin{figure}
\centering
\scalebox{0.33}{\includegraphics{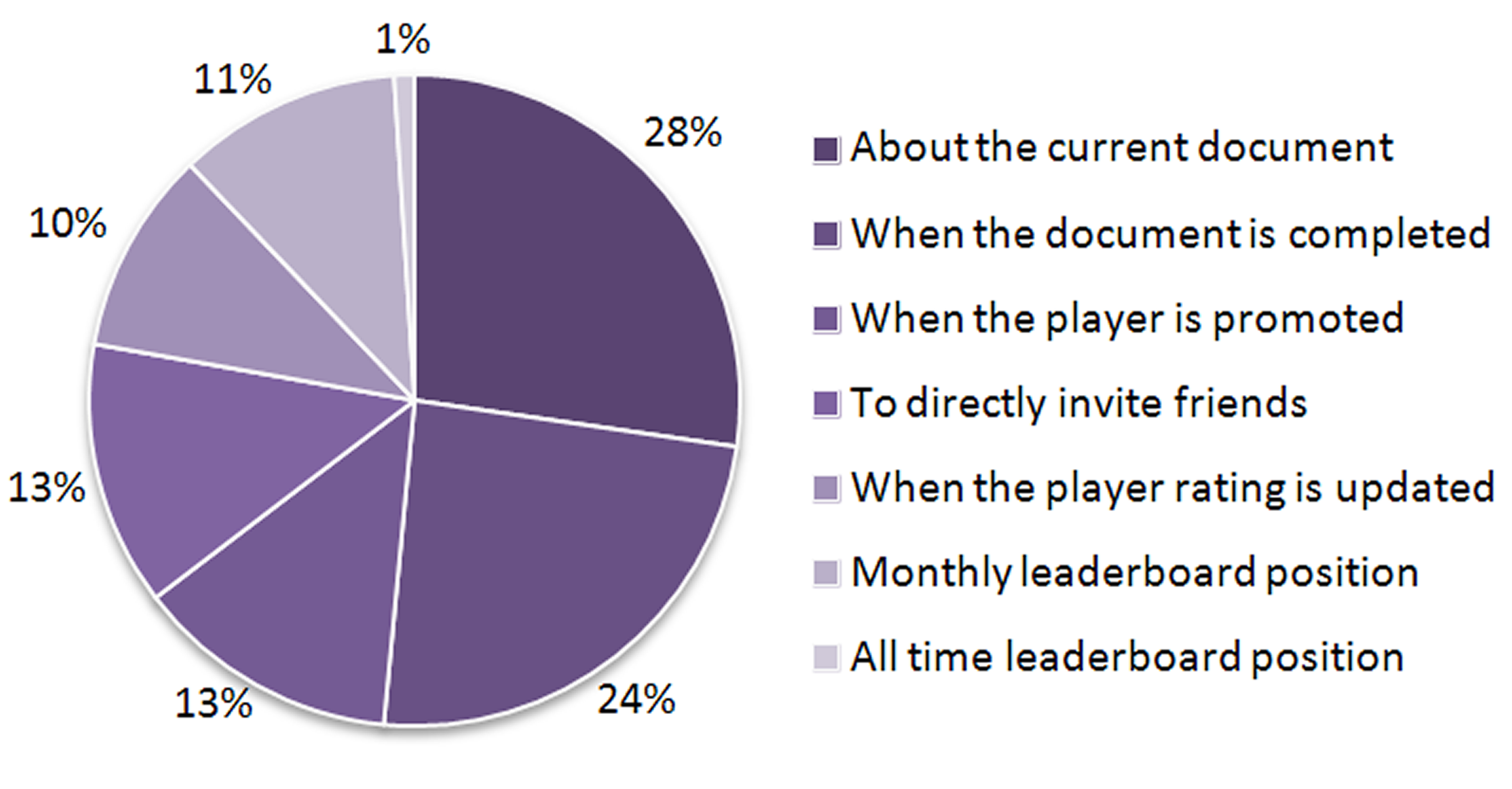}}
\caption{\label{R_wallposts} Chart showing the breakdown of 423 wall posts made by players.}
\end{figure}

Players' posts (see Figure \ref{R_wallposts}) were most commonly social i.e., about the document the player was working on or had just completed (52\%). This compares to competitive posts when they went up a level (13\%), when their rating was updated (10\%) or to post about their position in the leaderboard (12\%). The remaining 13\% of news posts were players making a direct collaborative request for their friends to join the game.

These results support the assumption that players are most likely to make posts from the game when the information is similar to what they might usually post. This should be a design consideration when developing a CI system for social networks. 

\begin{figure*}
\centering
\scalebox{0.55}{\includegraphics{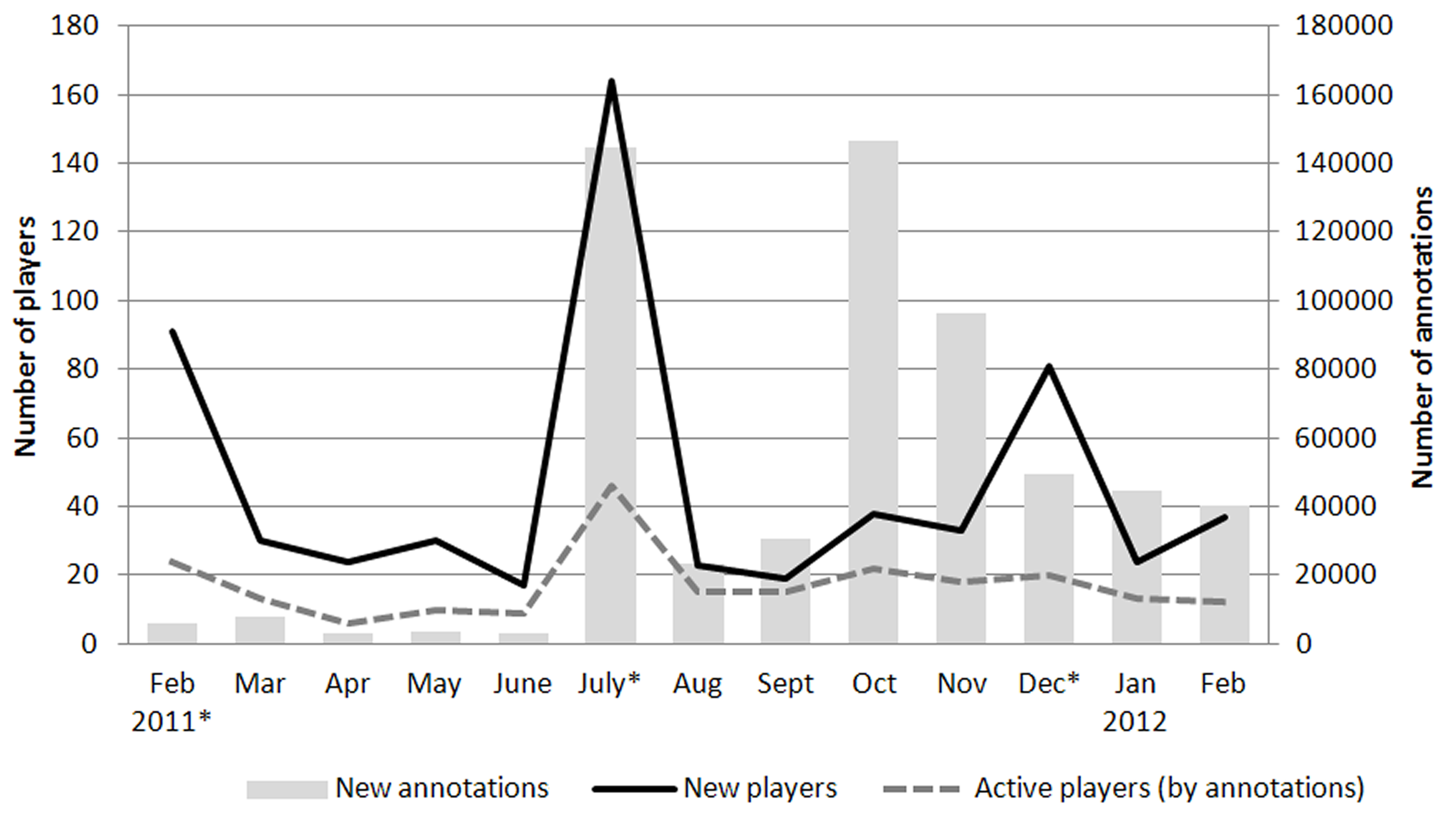}}
\caption{\label{R_promotion} Chart showing new annotations plotted against new players and active players. Prizes were available in the game from July 2011 to February 2012 (see Table \ref{table}).  * indicates a month with active promotion for the game.}
\end{figure*}

\subsubsection{The effect of incentives on recruitment and participation}

Figure \ref{R_promotion} shows the growth of {\it Phrase Detectives} on Facebook. Months where there was active promotion of the site (February, July and December 2011) show increases in new players, as one would expect. The month with the largest prize fund also had the most recruitment, double that of a later month with a smaller prize fund. 

Based on the assumption that the first promotion month, when the site went live, was an exception as players of the previous game joined the new version, there is an indication that financial incentives increase recuitment to the game if sufficiently advertised.

It is noticable that the amount of active players (a player who made more than one annotation in a particular month) stayed consistent and does not seem to increase with recruitment or financial incentives. Whilst it could be expected that the number of active players steadily increases over time as more players are recruited, the results show that most players will play the game for a short period of time and only small number continue to play every month. This is confirmed by the previous results that only a handful of players do most of the work.

Financial incentives do appear to be a strong motivating factor when considering how much work the active players do. Months with prizes have considerably more new annotations than those without, but with a similar number of active players. 

This suggests that active players are motivated by financial incentives, however the large amount of gameplay in October and November 2011 indictae that other motivating factors such as personal and social incentives are, to some extent, also successful.

\subsubsection{Cost effectiveness of financial incentives}

The cost per annotation is a measure of the effectiveness of the prize fund. A baseline of 4699 new annotations per month could be expected without a prize fund (the average of the first 5 months' new annotations) so anything over that could be attributed to the prize fund. 

The average cost per annotation across the months where there was a prize fund was \textsterling 0.0028 (see Table \ref{table}) and this can be compared to other CI systems. Amazon Mechanical Turk (MTurk)\footnote{http://www.mturk.com} is often used as a way of getting data quickly but can cost \textsterling 0.007 - 0.70 (\textdollar 0.01 - 0.10) per annotation \cite{Mason2010}. 

Both systems require the annotations to be aggregated before a collectively intelligent answer can be produced however even professional annotation schemes require some degree of validation. 

The advantage of GWAP over MTurk is that personal and social incentives can be used, as well as financial, to minimise the cost and maximise the persistence of the system.

It is also worth considering the setup and maintenance costs of CI systems in a cost per annotation analysis. 


\section{Conclusions}

In addition to accessing the vast human resources that social networking platforms offer, CI systems can also take advantage of the inherent social structure and shared personal data available to maximise the incentives that encourage participation.

{\it Phrase Detectives} has shown some valuable insights into user motivations in social network gaming and participation in CI efforts. The results support previous surveys that show women are more likely to play, and will spend more time playing, socially networked games. 


There are indications that players want to share information from the game that is similar to the information usually shared in their social networks. The success of a socially networked game relies on creating an experience that players want to share with their friends.

The results suggest that attracting and motivating the right kind of player is as important as attracting lots of players because, although collective intelligence needs a crowd, that crowd also needs to do some work. Financial incentives, coupled with promotion, increase recruitment and have a considerable impact on participation. 

The increase in data collected from the game due to financial incentives still makes it a cost effective alternative to other CI systems such as MTurk.

\section{Acknowledgments}
Thanks to Ans Alghamdi for his work on the Facebook version of the game.
The creation of the original game was funded by EPSRC project AnaWiki, EP/F00575X/1.

\begin{table}
\centering
\caption{\label{table} Cost per annotation based on the prize fund.}
\begin{tabular}{|l|r|r|} \hline
Month&Prize fund&Cost/annotation\\ \hline
Feb - Jun 2011&-&-\\ \hline
Jul 2011&320&0.0022\\ \hline
Aug 2011&110&0.0058\\ \hline
Sep 2011&110&0.0042\\ \hline
Oct 2011&110&0.0008\\ \hline
Nov 2011&110&0.0012\\ \hline
Dec 2011&110&0.0025\\ \hline
Jan 2012&110&0.0028\\ \hline
Feb 2012&110&0.0031\\ \hline
\hline\end{tabular}
\end{table}

\bibliography{coll_int}  

\end{document}